\documentclass[useAMS,usenatbib]{mn2e}

\usepackage{graphicx} \usepackage{natbib}

\newcommand{\logg}{\mbox{$\log g$}}

\def\m2s2{\hbox{\,m$^{2}$\,s$^{-2}$}} 
\def\Msun{\hbox{$\mathrm{M}_{\odot}$}} 
\def\Rsun{\hbox{$\mathrm{R}_{\odot}$}} 
\def\Mjup{\hbox{$\mathrm{M}_{\rm Jup}$}}
\def\Rjup{\hbox{$\mathrm{R}_{\rm Jup}$}}

\def\1s{$1\,\sigma$} 
 
\def \t0{T$_0$}

\usepackage{times}

\title[Transit observations of the exoplanet WASP-13b]{High precision transit observations of the exoplanet WASP-13b with the RISE instrument} \author[S.C.C. Barros et al.]{S. C.  C. Barros
  $^{1}$\thanks{E-mail:s.barros@qub.ac.uk}, D. L. Pollacco $^{1}$,
  N. P. Gibson $^{2}$, F. P. Keenan $^{1}$, I. Skillen$^{3}$,  I. A. Steele$^{4}$\\
  $^{1}$Astrophysics Research Centre, School of Mathematics and
  Physics, Queen's University Belfast, University Road, Belfast BT7
  1NN\\
  $^{2}$Department of Physics, University of Oxford, Denys Wilkinson
  Building, Keble Road, Oxford
  OX1 3RH \\
   $^{3}$Isaac Newton Group of Telescopes, Apartado de Correos 321,
  E-38700 Santa Cruz de la Palma, Tenerife,
  Spain\\
  $^{4}$Astrophysics Research Institute, Liverpool John Moores University, Wirral CH61 4UA}

\begin{document}

\date{Accepted . Received ; in original form }

\pagerange{\pageref{firstpage}--\pageref{lastpage}} \pubyear{2002}

\maketitle

\label{firstpage}

\begin{abstract}
  WASP-13b is a sub-Jupiter mass exoplanet orbiting a G1V type star
  with a period of $4.35\,$days.  The current uncertainty in its
  impact parameter ($0<b<0.46$) resulted in poorly defined stellar and
  planetary radii.  To better constrain the impact parameter we have
  obtained high precision transit observations with the RISE
  instrument mounted on $2.0\,$m Liverpool Telescope.  We present four
  new transits which are fitted with an MCMC routine to derive
  accurate system parameters.  We found an orbital inclination of
  $85.2 \pm 0.3$ degrees resulting in stellar and planetary radii of
  $1.56 \pm 0.04\,$\Rsun\ and $ 1.39 \pm 0.05\, $\Rjup, respectively.
  This suggests that the host star has evolved off the main-sequence
  and is in the shell hydrogen-burning phase. We also discuss how the
  limb darkening affects the derived system parameters. With a density
  of $0.17\, \rho_J$, WASP-13b joins the group of low density planets
  whose radii are too large to be explained by standard irradiation
  models.  We derive a new ephemeris for the system, $\mathrm{T}_0 =
  2455575.5136 \pm 0.0016$ (HJD) and $P=4.353011 \pm 0.000013\,$
  days. The planet equilibrium temperature ($T_{equ}=1500\,$K) and the
  bright host star ($V=10.4\,$mag) make it a good candidate for
  follow-up atmospheric studies.
\end{abstract}


\section{Introduction}

The discovery of the first short-period transiting planet HD 209458b
\citep{Charbonneau2000, Henry2000} opened a new field of exoplanetary
research. Their advantageous geometry allow us to estimate accurate
planetary mass and radii. The bulk density of the planet provides
information on its composition \citep{Guillot2005, Fortney2007} and
places constraints on planetary structure and formation models.

Interestingly, many of the short-period planets are bloated relative
to baseline models ( e.g. \citealt{Bodenheimer2003, Fortney2007}),
which assume a hydrogen/helium planet contracting under strong stellar
irradiation.  We define the radius anomaly of exoplanets as the
difference between the measured radius and theoretical radius.
Negative radius anomalies, i.e. denser planets, can be explained by
different planet composition or/and the presence of a core
\citep{Guillot2006,Burrows2007}. However, the majority of the hot
giant planets have low density (i.e. radius anomalies larger than
zero) that are harder to reconcile. Several theories have been
presented to explain the positive radius anomalies of hot exoplanets:
these include tidal heating \citep{Bodenheimer2001}, kinetic heating
due to winds \citep{Guillot2002}, enhanced atmospheric opacities
\citep{Burrows2007} and ohmic dissipation
\citep{Batygin2011}. Although these theories can explain individual
systems, only the recent ohmic dissipation theory \citep{Batygin2011},
can successfully explain all of the known exoplanets radius anomalies
\citep{Laughlin2011}. The lowest density hot Jupiters, such as
WASP-13b, are particularly important to shed light on the physical
mechanisms that lead to inflated planets.

For bright transiting systems, further insight into their physical
properties can be obtained through followup observations. Transmission
spectroscopy, which consists of measuring the stellar light filtered
through the planet's atmosphere during transit, provides information
into exoplanet atmospheres
\citep{Charbonneau2002,Vidal-Madjar2003}. Observation of secondary
eclipses (i.e. occultations) offers the potential for directly
measuring planetary emission spectra
(e.g. \citealt{Deming2005,Grillmair2008, Charbonneau2008}). Finally,
the ability to measure the sky-projected angle between the stellar
rotation axis and planetary orbit, through the Rossiter-McLaughlin
effect \citep{Rossiter1924,McLaughlin1924}, can yield clues about the
formation and migration processes
\citep{Fabrycky2009,Triaud2010,Winn2010}.

Followup opportunities benefit from accurate planetary and stellar
parameters.  However, obtaining high signal-to-noise transit
observations from the ground is difficult. Consequently even some of
the planets with bright host stars are lacking good quality light
curves, and hence have poorly determined planetary parameters. This is
the case for WASP-13b \citep{Skillen2009} where it was not possible to
accurately constrain the impact parameter ($0<b<0.46$) due to the lack
of a high precision transit light curve.

WASP-13b is a low-mass planet with $M_p = 0.46 \pm 0.06 $ \Mjup\, in a
4.3 day circular orbit \citep{Skillen2009}. Its host star is G1V type
with $M_*=1.03 \pm 0.10\,$\Msun\ , $T_{eff} = 5826 \pm 100\,$K, $\logg
= 4.04$ and solar metallicity. WASP-13b was discovered in 2006-2007 by
the SuperWASP-north survey \citep{Pollacco2006}. As mentioned above,
the impact parameter is not well constrained and two possible sets of
system parameters are presented in the discovery paper for two
possible impact parameters; $b=0$ and $b=0.46$. This results in large
uncertainties in both stellar ($R_*\sim 1.2-1.34 \Rsun$), and
planetary radii ($Rp \sim 1.06-1.21$). The adopted solution $b=0.46$
suggests that the host star has evolved off the main-sequence to the
shell hydrogen burning phase. Therefore, this system is interesting in
the context of exoplanet evolution, and with a bright host star
(V=10.4 mag) is a good candidate for followup studies.

In this paper, we present four high precision transit observations of
WASP-13b which are described in Section 2.  Our data allow us
  to determine the inclination of the system and directly derive the
  mean stellar density and hence, better constraint the stellar and
  planetary radii. We discuss our transit model in Section 3 and
present the updated parameters of the system in Section 4. In Section
5 we discuss the discrepancy between theoretical and empirical limb
darkening coefficients. Finally, we discuss our results in Section 6.

\section{Observations}

WASP-13b was observed with RISE
\citep{rise2008,Gibson2008,Barros2011b} mounted on the 2.0m Liverpool
Telescope on La Palma, Canary Islands. RISE is a frame transfer e2v
CCD with a pixel scale of 0.54 arcsec/pixel that results in a 9.4
$\times$ 9.4 arcmin field-of-view. It has a wideband filter covering
$\sim 500$--$700\,$nm which corresponds approximately to V+R.  For
exposures longer than 1 second, RISE has a deadtime of only 35 ms.  To
reduce the deadtime to the minimum \citep{Barros2011b} we used an
exposure time of 1 second which required a defocussing of -1.0 mm to
avoid saturation of the bright comparison star
($8.7\,$mag). Defocussing is also important to decrease systematic
noise due to poor guiding \citep{Barros2011b}.

The data were reduced using the \textsc{ultracam} pipeline
\citep{Ultracam} which is optimized for time-series photometry. Each
frame was bias subtracted and flat field corrected. We performed
differential photometry relative to an ensemble of comparison stars in
the field, confirmed to be non-variable. For each observation we
sampled different aperture radii and chose the aperture radius that
minimised the noise. The photometric errors include shot noise,
readout and background noise.

On 2009-01-29 we obtained an ingress of WASP-13b and an egress on
2009-05-05. We obtained full transits of WASP-13b on 2010-02-03 and
2011-01-13. Unfortunately, both observations were degraded by some
clouds. Observations with transparency lower than 94\% were
  clipped. On 2010-02-03 the transit observation was interrupted
after ingress due to the derotator reaching its limiting position and
the two parts of this transit are treated as separate
observations. This observation shows systematics near the
  rotator interruption and is affected by clouds before the egress. We
  experimented removing it from the final analysis and concluded that
  its inclusion does not bias the results. On the last 20 minutes of
  the 2011-01-13 observation the position of the star in the CCD
  varied by 10 pixels in the x and 15 pixels in y which caused an
  additional trend in the light curve. Due to its effect on the
  normalisation of the light curve, the last 20 minutes of the
  2011-01-13 observation were clipped. The number of exposures
taken, estimated FWHM and aperture radius used for each observation
are given in Table~\ref{obser}. We also included in our analysis, the
previously published transit of WASP-13 taken with $0.95\,$m James
Gregory Telescope (JGT) at St Andrews University on 2008-02-16
\citep{Skillen2009}.

\begin{table}[h]
  \centering 
  \caption{WASP-13 observation log.}
  \label{obser}
  \begin{tabular}{lccccl}
    \hline
    \hline
    Date & number exposures & FWHM & aperture radius  \\
    &  & pixel & pixel  \\
    \hline
    2009-01-29 & 4400  & 17 & 23 \\
    2009-05-05 & 10780 & 21 & 24 \\
    2010-02-03 & 17000 & 17 & 22 \\
    2011-01-13 & 16000 & 24 & 27 \\
    \hline
  \end{tabular}
\end{table}

\begin{figure}
  \centering
  \includegraphics[width=\columnwidth]{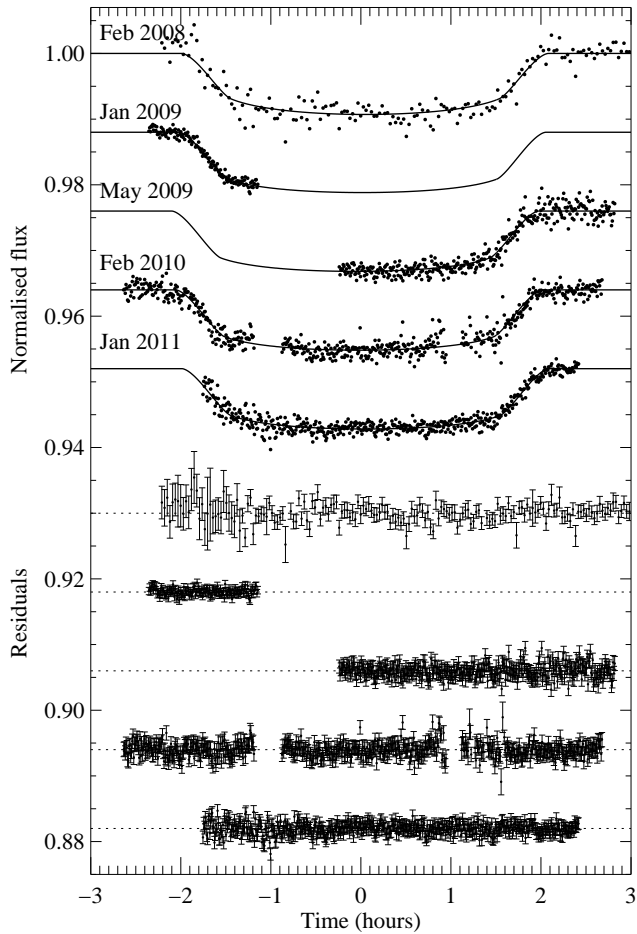}
  \caption{Transit observations of WASP-13b. From top to bottom in
    chronological order; JGT 2008 February 16, RISE 2009 January 01,
    RISE 2009 May 05, RISE 2010 February 03 and RISE 2011 January
    13. We superimpose the best-fit transit model and also show the
    residuals for each light curve at the bottom of the figure. The
    RISE data are binned into 30 second periods, and bins displaced
    vertically for clarity. The individual RISE light curves plotted
    here are available in electronic form at CDS.}
  \label{photolc}
\end{figure}

\begin{figure}
  \centering
  \includegraphics[width=\columnwidth]{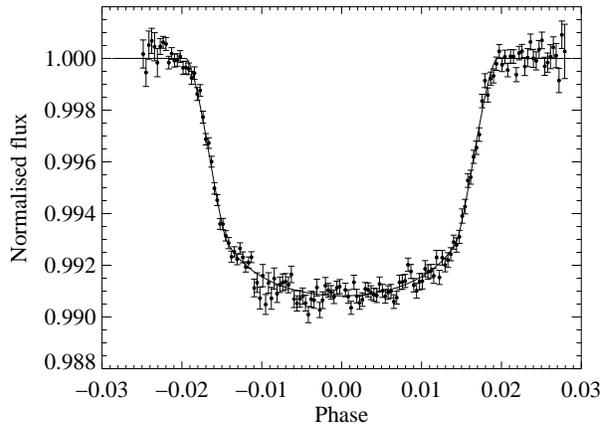}
  \caption{Phase-folded combined transit light curve of WASP-13b.  The
    data was optimally binned in 150 phase bins for clarity. We
    superimpose the best-fit transit model. Note that here we do not
    account for the different limb darkening of the RISE and the JGT
    light curves.}
  \label{phase}
\end{figure}

The final high precision transit light curves for WASP-13b are shown
in Figure~\ref{photolc} along with the best-fit model described in
Section~\ref{model}. We overplot the model residuals and the estimated
uncertainties, which are discussed in Section~\ref{errors}.  To
illustrate the quality of our results we also show the phase folded
weighted combination of all the light curves in Figure~\ref{phase}.
Around -0.01 in phase the only high precision data is from the
  2011-01-13 light curve that shows extra scatter.  is below the
  transit model. The remaining transit phases are better sampled and
  hence have smaller errors.

\section{Data analysis}

\subsection{Photometric errors}
\label{errors}

As mentioned above the initial photometric errors include only shot
noise, readout and background noises, which underestimates the true
errors. To obtain a more reliable estimate of the errors we begin by
scaling the errors of each light curve so that the reduced $\chi^2$ of
the best fitting model is 1.0. This resulted in the multiplication of
the errors by $1.01$, $1.55$, $2.46$, $2.68$, $2.26$ and $2.79$, for
the 2008, Jan 2009, May 2009, ingress 2010, egress 2010 and 2011 light
curves, respectively. For the RISE observations, this indicates that
the photometric errors are higher than expected from photon noise,
probably due to the non-photometric conditions of these observations.

Exoplanet transit observations are also affected by time correlated
noise which can lead to underestimated parameter uncertainties
\citep{Pont2006,carter2009}. The residuals displayed in
Figure~\ref{photolc} show that the photometric noise is dominated by
white noise. Nevertheless, we included time-correlated noise following
the procedure from \citet{gillon2009}. Using the residuals of the best
fit model, we estimated the amplitude of the red noise, $\sigma_r$, to
be $100\,$ppm (Jan 2009), $50\,$ppm (May 2009), $250\,$ppm (ingress
2010), $150\,$ppm (egress 2010), and $200\,$ppm (2011). For the 2008
light curve, the red noise was found to be negligible. These errors
were added in quadrature to the rescaled photometric errors and were
used in the final Markov Chain Monte Carlo (MCMC) chains.

\subsection{Determination of system parameters}
\label{model}
To determine the planetary and orbital parameters, we fitted the four
RISE light curves and the original JGT light curve of WASP-13b
simultaneously. We used the same procedure as for WASP-21b
\citep{Barros2011b} which is based on the \citet{Mandel2002} transit
model parametrised by the normalised separation of the planet,
$a/R_*$, ratio of planet radius to star radius, $ R_p/R_* $, orbital
inclination, $i$, and the transit epoch, $T_0$, of each light
curve. We account for a linear trend with time for each light curve
parametrised by the out of transit flux and flux gradient. Due to the
short orbital period and age of the system we expect the orbit has
circularised (see below), hence a circular orbit was assumed. The
transit model is coupled with an MCMC procedure to obtain accurate
parameters and uncertainties as outlined in \citet{Barros2011b}. We
included the quadratic limb darkening coefficients (LDC) from the
models of \citet{Howarth2011a} for $T_{eff} = 6000\,$K, $\logg=4.0$
and $[M/H] = 0.0$ which were the closest tabulations to the parameters
from \citet{Skillen2009}; for the RISE filter V+R, $a=0.4402$ and
$b=0.2394$ and for the R filter of JGT, $a=0.4121 $ and $b=0.2312$. We
initially kept the limb darkening parameters fixed during the fit. We
treated the two parts of the 2010-02-03 observations as separate
transits to correctly account for the normalisation but restricted
them to have the same $T_0$. Including the two linear normalisation
parameters for each light curve, a total of 20 parameters were
fitted. Besides the linear normalization, no extra trends were removed
from the light curves.

We computed seven MCMC chains, each of $500\,000$ points and different
initial parameters. The initial 20\% of each chain that corresponded
to the burn in phase were discarded and the remaining parts merged
into a master chain. We define the best fit parameter as the
  mode of its probability distribution and the 1 $\sigma$ limits as
  the value at which the integral of the distribution equals 34.1\%
  from each side of the mode. The \citet{Gelman92} statistic was
calculated for each fitted parameter and we concluded that chain
convergence was good.

To test how the limb darkening coefficients affect the derived system
parameters we experimented with fitting the limb darkening
coefficients of the RISE light curves which have higher precision.
Our tests indicate that the quality of the RISE light curves is not
enough to fit for both limb darkening coefficients (see also
\citealt{Gibson2008}), and hence, we choose to fit the linear LDC $a$
which dominates. The quadratic LDC of the RISE light curves $b$ and
both LDCs of the JGT light curve were kept fixed. Therefore, in this
second MCMC procedure 21 parameters were fitted.

\section{Results}

In Table~\ref{mcmc}, we present the system parameters of WASP-13 and
the 1$\sigma$ uncertainties derived from the MCMC analysis for both
the fixed and fitted LDCs. The first set of parameters can be directly
derived from the light curve analysis. The parameters were derived
using the equations of \citet{Seager2003, Southworth2007,
  Kipping2010}. Specifically, the $T_{1-4}$ refers to the total
transit duration and $T_{T1}$ is defined by equation 15 of
\citet{Kipping2010}.

The derived linear limb darkening coefficient, $a = 0.34 \pm 0.03$, is
statistically significantly different from the theoretical value
$a=0.4402$. The chain convergence was good and the linear LDC
parameter is well constrained by our observations. We adopt the fitted
LDC solution because according to the F-test it is significantly
($99.9\%$) better than the fixed LDC one. In section 5 we present a
more in depth discussion about the discrepancy between the fitted and
theoretical LDCs.

All the parameters of both solutions are within $\sim 3\sigma$.  We
found that the inclination of the orbit is $85.2 \pm 0.3\,$ degrees
and the derived stellar density is $ \rho_* = 0.29 \pm 0.02
\rho_{\odot}$. This is significantly lower than previously assumed ($
\rho_* = 0.43 \rho_{\odot}$ \citealt{Skillen2009}), implying an
evolved star. To obtain the stellar and planetary radii we need to
determine the stellar mass, as explained in the next section.

\begin{table}[h]
  \centering 
  \caption{WASP-13 system parameters.}
  \label{mcmc}
  \begin{tabular}{lcccl}
    \hline
    \hline
    Parameter & LDC fixed & LDC fitted  \\
    \hline
    Linear LDC for RISE $a$ & $0.4255$ & $0.337 \pm 0.033$ \\
    Normalised separation $a/R_*$ & $7.58^{+0.15}_{-0.13}$  & $7.39^{+0.14}_{-0.15}$   \\
    Planet/star radius ratio $ R_p/R_* $ & $ 0.09225 332^{+0.00080}_{-0.00081}$& $ 0.09226^{+0.00083}_{-0.0010}$  \\
    Transit duration $T_{1-4}$ [hours] & $ 4.094^{0.024}_{-0.033}$ & $ 4.063^{0.026}_{-0.032}$ \\
    Transit duration $T_{T1}$ [hours] & $ 3.607^{0.012}_{-0.015}$ &  $ 3.548^{0.019}_{-0.017}$\\
    Orbital inclination $ I $ [degrees] & $  85.64 \pm 0.24$ & $  85.19 \pm 0.26$ \\    
    Impact parameter $ b $ [$R_*$]  & $ 0.575 \pm 0.020$  & $ 0.621 \pm 0.021$  \\      
    Stellar density $ \rho_* $ [$\rho_{\odot}$] & $ 0.309^{+0.018}_{-0.016}  $ & $ 0.288^{+0.019}_{-0.015}  $ \\
    &    &      &  \\
    Orbital semimajor axis $ a $ [AU] & $  0.05379^{+0.00077}_{-0.00059}$ & $  0.05362^{+0.00074}_{-0.00085}$ \\
    Stellar mass $ \mathrm{M}_* $ [\Msun] & $ 1.085 \pm 0.04   $ & $ 1.09 \pm 0.05   $  \\
    Stellar radius $ \mathrm{R}_* $ [\Rsun]  & $   1.512^{+0.031}_{-0.041}$& $   1.559 \pm 0.041$ \\
    &    &      &  \\
    Planet mass  $ \mathrm{M}_p $ [\Mjup]  & $ 0.485^{+0.042}_{-0.058} $ & $ 0.477^{+0.044}_{-0.049} $  \\
    Planet radius $ \mathrm{R}_p $ [\Rjup]  &  $1.365^{+0.034}_{-0.062}$  &  $1.389^{+0.045}_{-0.056}$ \\
    Planet density  $ \rho_p $ [$\rho_J$]  &  $0.190^{+0.028}_{-0.014}$ &  $0.167^{+0.017}_{-0.022}$\\
    \hline
  \end{tabular}
\end{table}

\subsection{Stellar mass and age}

\begin{figure}
  \centering
  \includegraphics[width=\columnwidth]{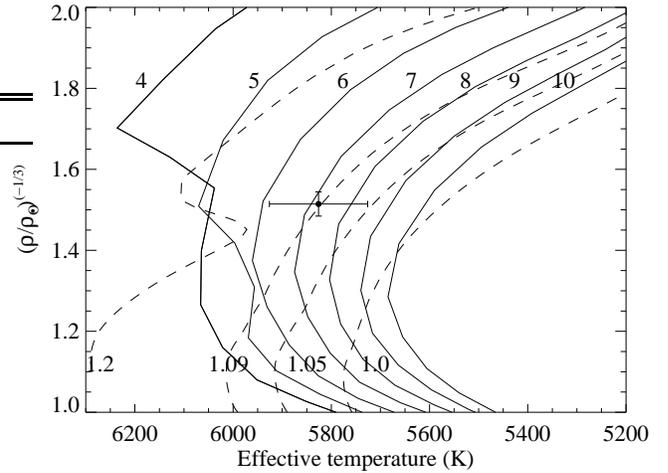}
  \caption{ Isochrone models (solid lines) from \citet{Demarque2004}
    for WASP-13 using {[Fe/H]} =0 from \citet{Skillen2009} and
    assuming solar composition Z=0.0181. The age in Gyr is marked in
    the left of the respective model. We also show the mass tracks
    (dashed lines) for $1.0, 1.05, 1.09$ and $1.2 \Msun$.  We overplot
    the $T_{eff} = 5826$ \citep{Skillen2009} and our fitted $( \rho_*/
    \rho_{\odot})^{-1/3} $ for WASP-13.}
  \label{isocrones}
\end{figure}

Our high quality transit light curves allow an accurate estimate of
the stellar density. This, combined with stellar models, can be used
to constrain the stellar mass and age.  We interpolate the Yonsei-Yale
evolution tracks \citep{Demarque2004} for $[Fe/H] =0$
\citep{Skillen2009} which are plotted in Figure~\ref{isocrones}.  Also
shown are the position of WASP-13 with the effective temperature from
\citet{Skillen2009} and our value of $\rho_* $.  We derive a stellar
mass of $1.09 \pm 0.05 \Msun$ and a stellar age of $7.4 \pm
0.4\,$Gyr. Both agree with the estimate in the original discovery
paper. Using the stellar reflex velocity from \citet{Skillen2009},
$K_1=55.7\,$ m/s, we estimate a planetary mass of $0.48\, \Mjup$ .

\begin{figure}
  \centering
  \includegraphics[width=\columnwidth]{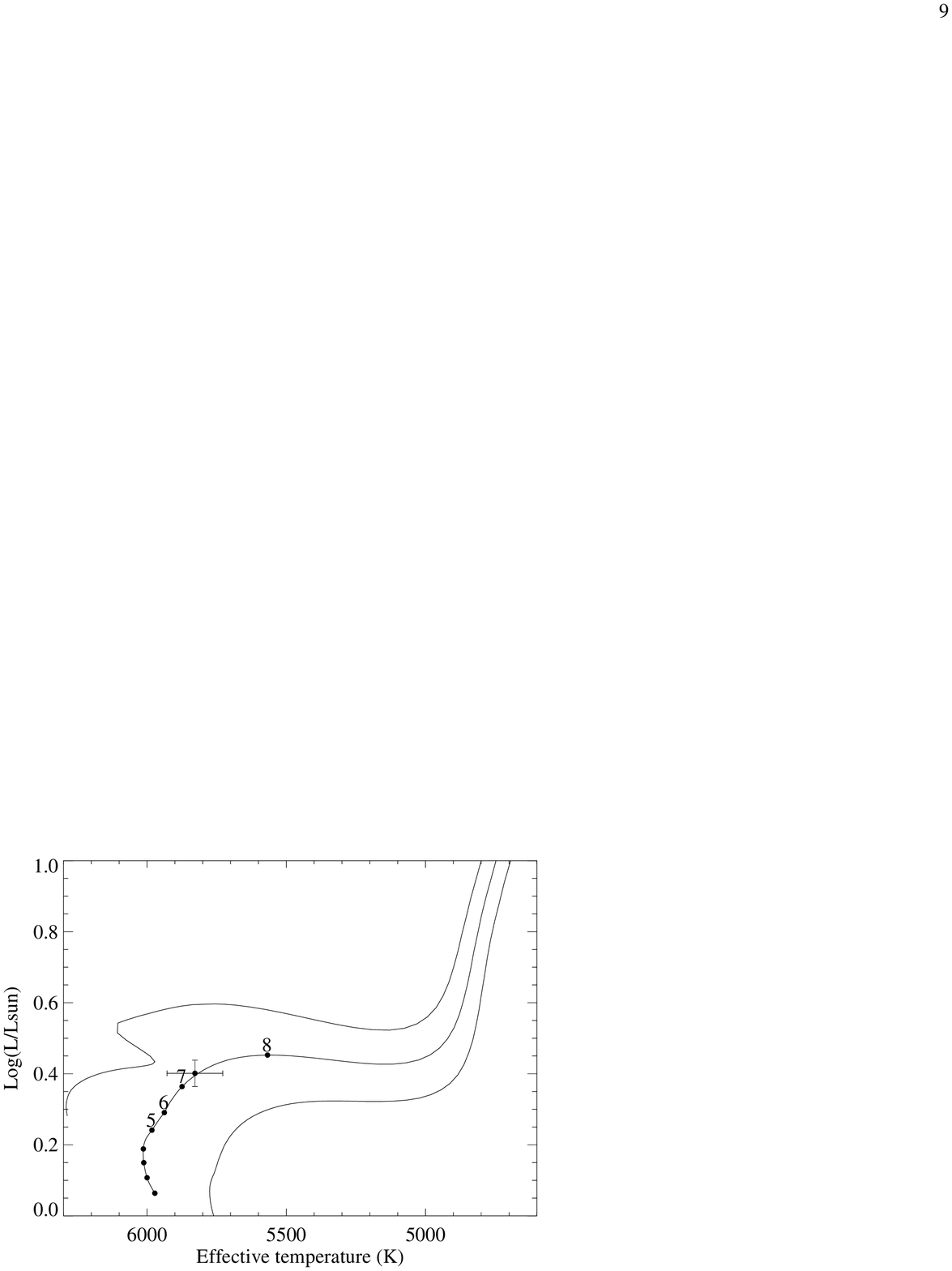}
  \caption{Evolutionary mass tracks from \citet{Demarque2004} for the
    same stellar parameters as Fig.~\ref{isocrones} for stellar masses
    of 1.2, 1.09 and 1.0 \Msun\ from left to right. For the 1.09
    \Msun\ evolutionary mass track we also show the 5, 6,7 and 8 Gyr
    points.}
  \label{masstrack}
\end{figure}

In Figure \ref{masstrack} we show the evolutionary tracks for a
stellar mass of 1.0, 1.09 and 1.2 \Msun\ adapted from
\citet{Demarque2004} and we overplot the position of WASP-13. These
models suggest that WASP-13 has evolved to the hydrogen-shell burning
phase and is close to entering the sub-giant branch.  However,
  the uncertainty in the WASP-13 metallicity ( $[Fe/H] =0 \pm 0.2$)
  affects the determination of the evolutionary status of WASP-13.
  According to the \citet{Demarque2004} models, if WASP-13 has a
  metallicity of +0.2 then it could have a mass of 1.2 \Msun which is
  above the critical mass for the stellar cores to become convective
  and where details on the treatment of convective overshooting became
  important. In this case, the evolutionary status would be more
  uncertain but probably WASP-13 would be in the contraction phase
  before the hydrogen-shell burning phase starts.

We obtain a larger stellar and planetary radius ($ \mathrm{R}_*=1.56
\pm 0.04\, \Rsun$ and $ \mathrm{R}_p= 1.39 \pm 0.05\, \Rjup$) than
previously reported in \citet{Skillen2009}. Our derived parameters are
consistent with the ``b=0.46'' adopted solution presented in the
discovery paper, although our precision is almost $10$ times better.

\subsection{Transit times}
The estimated transit times were used to update the linear ephemeris,
\begin{equation}
  T (HJD) = T_0 + EP.
\end{equation}
We found $P=4.353011 \pm 0.000013$ days and T$_0=2455575.5136 \pm
0.0016$, which was set to the mid-transit time of the 2011 light
curve. The updated period was used in the final MCMC procedures.

Time residuals from the linear ephemeris are given in
Table~\ref{ttv}. The partial transits show a larger deviation from the
linear ephemeris which was also noted for the case of TrES-3b
\citep{Gibson2009}. These authors concluded that the uncertainties in
the transit times of partial transits are underestimated, probably due
to the lack of out-of-transit data needed for
normalisation. Therefore, although the time residual of the May2009
transit is significant at 4 sigma, this could be due to normalisation
errors rather than an intrinsic transit timing
variation. 
For example, removing the out-of-transit data of the May2009 light
curve results in a difference in the estimated mid-transit time of
$155\,$s $=3.5\sigma$. This suggests the presence of additional red
noise in the light curve that affects the transit times which should
be regarded with caution.  Hence, we conclude that the time residuals
of WASP-13b are probably consistent with a linear ephemeris, but
encourage further transit timing observations.

\begin{table}[h]
  \centering 
  \caption{Time residuals from the linear ephemeris.}
  \label{ttv}
  \begin{tabular}{cccc}
    \hline
    \hline
    Date &    Epoch & Time residuals (sec) & Uncertainty (sec) \\
    \hline
    2008-02-16 & -244 &     235&           82\\
    2009-01-29 & -164 &      83&           42\\
    2009-05-05 & -142 &    -211&           51\\
    2010-02-03 &  -79 &     -92&           39\\
    2011-01-13 &   0  &     114&           44\\
    \hline
  \end{tabular}
\end{table}

\section{Limb darkening}
Our measured linear limb darkening parameter ($a = 0.34 \pm 0.03$) is
statistically different from the one predicted from stellar models
($a=0.4402$). We choose the theoretical LDCs from stellar models of
\citet{Howarth2011a} because they include the 'V+R' RISE filter.

Disagreement between the empirical and theoretical LDCs has been
reported for other systems (e.g. \citealt{Southworth2008,
  Barros2011b}) including for higher quality light curves from HST
\citep{Claret2009}, CoRoT (\citealt{sing2010}, Csizmadia, Sz. et al in
prep) and Kepler \citep{Kipping2011}. This suggests that the
discrepancy is not due to insufficient quality of our light curves.
Moreover, \citet{Claret2009} concluded that uncertainties in the
stellar temperature and metallicity cannot explain the difference
between the theoretical and empirical LDCs. Therefore, the causes of
this discrepancy are not clear.

Recently, \citet{Howarth2011b} showed that LDCs derived from transit
observations depend on the impact parameter and cannot be directly
compared with the theoretical values. This is because for a higher
impact parameter the transit shape is less sensitive to the global LD
law. This can lead to up to a 60\% difference in the estimated LDCs
\citep{Howarth2011b}. Fortunately, this insensitivity also means that
the effect on the transit parameters is relatively small. Using a
synthetic photometry/atmosphere-model, \citet{Howarth2011b} predicts
that this geometric effect leads to an underestimation of the linear
LDC of the quadratic law which qualitatively explains our estimated
linear LDC. If this was the only cause of the discrepancy we should
adopt the theoretical LDC solution for the system parameters because
it would remove the correlation between the LDCs and the impact
parameter providing more accurate results.

Using HST and Kepler data, \citet{Howarth2011b} concluded that
accounting for this geometrical dependency leads to better agreement
between the measured and predicted LDCs. However, this cannot
reproduce all of the empirical data suggesting that there could be
other causes of the discrepancy. Tentative explanations include the
missing physics in the atmospheric models like, for example, the
effects of faculae and granulation (Csizmadia, Sz. et al in prep.) and
the plane-parallel approximation in stellar models
\citep{Neilson2011}. In this case, it would be advisable to adopt the
fitted LDC solution for the system parameters because it is less
dependent on the stellar models.

Further investigation of very high precision transit light curves such
as from the Kepler mission is needed to better understand the
difference between the measured and theoretical LDCs. This will help
to decide if we should fit for the LDCs in lower quality light curves.

\section{Discussion and Conclusion}
We have presented four new transit light curves of WASP-13b including
two full transits. To derive the parameters of the system, we used the
same MCMC procedure as \citet{Barros2011b} by fitting these new light
curves together with a previous JGT transit of WASP-13b
\citep{Skillen2009}. We estimated the uncertainties accounting for
formal errors and systematic noise.

Our derived value of the linear LDC is statistically significantly
different from the theoretical value predicted for stellar atmospheres
models. Although this does not significantly affect the derived system
parameters it can be important for stellar atmosphere models. We adopt
the fitted LDC solution because it is a statistical significantly
better fit to the data.

The impact parameter of WASP-13b was poorly constrained in the
discovery paper ($0<b<0.46$) though the higher value was adopted. We
obtain $b=0.62 \pm 0.02$, which is close to the previous adopted value
although slightly higher. This implies a larger stellar radius of $
R_*=1.56 \pm 0.04 \Rsun$ which combined with the evolution models of
\citet{Demarque2004} points to an age of $7.4 \pm 0.4\,$Gyrs for the
host star. It also suggests that WASP-13 has evolved off the
  main-sequence and is now in the shell hydrogen burning phase which
  would imply that in less than $2\,$Gyr the star will engulf the
  planet's current orbit. The evolutionary status of WASP-13 favours
a slightly higher mass for the host star of $\sim 1.1\Msun$ which also
implies a slightly higher planetary mass of $ 0.48 \Mjup$.

We derived a larger planetary radius for WASP-13b, ($ R_p= 1.39 \pm
0.05 \Rjup$), than previously reported and hence a lower density, ($
\rho_p = 0.17 \pm 0.01 \rho_J$). Therefore, WASP-13b is the fifth
lowest density exoplanet known after Kepler-7b \citep{Latham2010},
WASP-17b \citep{Anderson2010}, TrES-4b \citep{Mandushev2007} and
CoRoT-5b \citep{Rauer2009} and its properties are important for
irradiation models.  We estimate an equilibrium temperature for
WASP-13b of $\sim 1500\,$K.  The hydrogen/helium coreless models of
\citet{Fortney2007} predict a radius of $\sim 1.1 \Rjup$ for
WASP-13b. This is in agreement with the radius derived from the
polynomial fit of \citet{Laughlin2011} to the baseline models of
\citet{Bodenheimer2003}, R$_p \sim 1.16$. We derive a radius anomaly
for WASP-13b of $\Re=0.23$ \citep{Laughlin2011}. Therefore, WASP-13b
is inflated relative to the theoretical thermal irradiation models for
gas giant planets.

The circularization timescale for WASP-13b is 0.011-0.11 Gyr for $Q_p
= 10^5-10^6$. Therefore, we expect a circular orbit and it is unlikely
that the inflated radius of WASP-13 could be explained by tidal
dissipation. Enhanced atmospheric opacities of $10 \times$ solar, plus
the correction due to the transit radius effect \citep{Burrows2007},
can explain a radius anomaly of $\sim 0.18$ for a 7.4 Gyr planet and
hence might be able to explain the radius anomaly of WASP-13b. The
models of \citet{Bodenheimer2003} which include kinetic heating
\citep{Guillot2002} can also explain the radius of WASP-13b. Finally,
the ohmic dissipation model of \citet{Batygin2011} explains all known
exoplanet radii.

According to the classification of \citet{Fortney2008}, WASP-13b
belongs to the highly irradiated ``pM'' class planets. These have
considerable opacities due to molecular hazes of TiO and VO and are
expected to have temperature inversions in their atmosphere. This can
be tested by secondary eclipse observations. The large scale height of
WASP-13b, $\sim 550\,$Km, and its bright host star, $V=10.4\,$, makes
it also a good target for transmission spectroscopy. Probing the
atmospheric composition of this planet can shed light on its inflated
radius.

\section{Acknowledgements}

The RISE instrument mounted at the Liverpool Telescope was designed
and built with resources made available from Queen's University
Belfast, Liverpool John Moores University and the University of
Manchester. The Liverpool Telescope is operated on the island of La
Palma by Liverpool John Moores University in the Spanish Observatorio
del Roque de los Muchachos of the Instituto de Astrofisica de Canarias
with financial support from the UK Science and Technology Facilities
Council.  We thank Tom Marsh for the use of the \textsc{ultracam}
pipeline.  We thank the anonymous referee for interesting suggestions.

\bibliographystyle{mn2e} \bibliography{susana_mn}

\label{lastpage}

\end{document}